# Shell feature: a new radiomics descriptor for predicting distant failure after radiotherapy in non-small cell lung cancer and cervix cancer

Hongxia Hao[1,2,3], Zhiguo Zhou[3], Shulong Li[3,4], Genevieve Maquilan[3], Michael R. Folkert[3], Puneeth Iyengar[3], Kenneth D. Westover[3], Kevin Albuquerque[3], Fang Liu[1,2], Hak Choy[3], Robert Timmerman[3], Lin Yang[5], Jing Wang[3]

1. School of Computer Science and Technology, Xidian University, Xi'an, 710071, China, 2. Key Laboratory of Intelligent Perception and Image Understanding of Ministry of Education, International Research Center for Intelligent Perception and Computation, Joint International Research Laboratory of Intelligent Perception and Computation, Xidian University, Xi'an, 710071, China, 3. Department of Radiation Oncology, University of Texas Southwestern Medical Center, Dallas, TX, 75235, United States, 4. School of Biomedical Engineering, Southern medical university, Guangzhou, 510515, China, 5. Department of Pathology, National Cancer Center/Cancer Hospital, Chinese Academy of Medical Sciences, Beijing, 100021, China.

**Acknowledgement:** This work was supported in part by the American Cancer Society (ACS-IRG-02-196) and US National Institutes of Health (5P30CA142543). The authors would like to thank Dr. Damiana Chiavolini for providing helpful suggestions and editing the manuscript.

**Corresponding Author:** Jing Wang, Associate Professor, Department of Radiation Oncology, UT Southwestern Medical Center, 2280 Inwood Rd. Dallas, TX, 75235-9303; e-mail: Jing.Wang@utsouthwestern.edu; Tel: 214-648-1795; fax: 214-645-2885.

**Running head:** Shell Feature for Predicting Distant Failure in NSCLC and CC

**Disclaimers:** The authors declare that they have no conflict of interest.

# ABSTRACT

**Purpose**
To develop and demonstrate a novel tumor shell feature for predicting distant failure in non-small cell lung cancer (NSCLC) and cervical cancer (CC) patients.
**Patients and Methods**
The shell predictive model was constructed using pre-treatment positron emission tomography (PET) images from 48 NSCLC patients received stereotactic body radiation therapy (SBRT) and 52 CC patients underwent external beam radiation therapy and concurrent chemotherapy followed with high-dose-rate intracavitary brachytherapy. A shell feature, consisting of outer voxels around the tumor boundary, was extracted from a series of axial PET slices. The hypothesis behind this feature is that non-invasive and invasive tumors may have different morphologic patterns in the tumor periphery, in turn reflecting the differences in radiological presentations in the PET images. The shell's utility was evaluated by the support vector machine (SVM) classifier in comparison with intensity, geometry, gray level co-occurrence matrix (GLCM)-based texture, neighborhood gray tone difference matrix (NGTDM)-based texture, and a combination of these four features. The results were assessed in terms of accuracy, sensitivity, specificity, and the area under the receiver operating curve (AUC).
**Results**
For NSCLC, the AUC achieved by the shell feature was 0.82 while the highest AUC achieved by the other features was 0.76. Similarly, for CC, the AUC achieved by the shell feature was 0.83 while the highest AUC achieved by the other features was 0.76. Also, the difference in performance between shell and the other features was significant ($P < 0.005$) in all cases.
**Conclusions**
We propose a boundary-based shell feature that correlates with tumor metastasis. The shell feature showed better predictive performance than all the other features for distant failure prediction in both NSCLC and CC.

## INTRODUCTION

Distant failure occurs when malignant tumor cells metastasize to distant organs,[1,2] causing up to 90% human cancers-associated deaths.[3,4] Stereotactic body radiation therapy (SBRT) is widely used in patients with early stage medically inoperable non-small cell lung cancer (NSCLC), achieving 85-95% local control rates[5-7]. Despite the high local control rates, distant failure is still common, with 3-year and 5-year distant relapse rates of 22% and 31%, respectively.[8-10] Similarly, in patients with locally advanced cervical cancer (CC), even receiving external beam radiation therapy (EBRT) with concurrent chemotherapy and intracavitary brachytherapy (ICBT) as the recommended therapy,[11] at least 20% of them still develop distant metastases;[12,13] and in the patients with positive para-aortic involvement, the rate is more than 40%.[14,15] Therefore, predicting distant failure in high risk patients is essential to achieve better treatment outcomes with intensified treatment modalities.

Although many of the mechanisms that govern metastasis are still unclear, the tumor microenvironment is known to regulate tumor evolution toward metastasis,[16,17] as shown for cervix,[18,19] lung,[20,21] colon cancer among other cancer types.[22,23]

A correlation was found between the microenvironment and distant failure,[24] typically exemplified by the theory of epithelial-to-mesenchymal transition (EMT). In this process, a portion of cancer cells located at the tumor edges may acquire cancer stem cell (CSC)-like traits typical of metastasis, including self-renewal, tumor-originating, invasiveness, and elevated apoptosis resistance; these cancer cells depart from the main tumor and initiate metastasis, leading to junctional alterations and spatial heterogeneity.[25-28] This cellular invasion process was simulated by a hybrid multiscale mathematical model, showing that invasive tumor cells first developed within the tumor and later penetrated the tumor edge to form metastases.[29]

In addition to EMT and CSCs, tumor budding[30] is another factor contributing to invasion and correlating with worse outcomes in colon cancer,[22,31,32] lung cancer,[33-36] cervix cancer[37] among others.[38-41] In tumor budding, isolated or clustered small malignant cells are close to the tumor edge. Literature reviews have reported that tumor buds can be a realization of CSC and an exhibition of the EMT process,[42,43] suggesting tumor budding as an independent prognostic factor.[37,44,45]

Tumor islands were also observed on tumor edges.[46] In lung cancer, tumor islands are large nests of malignant cells connected with one another and with primary tumors in alveolar spaces, slightly near the tumor border; they also have been associated with poor prognosis.[47] Studies were expanded by discovering spread through air space (STAS), a phenomenon of aggressive cells within air spaces closely beyond the edge of the tumor. STAS has been recognized an important pattern of invasion, [48-50] and was approved by the 2015 World Health Organization as an independent metastatic predictor of lung cancer within the lung classification system. [51]

These findings suggest that the appearance of the interface between tumor and normal tissue may provide phenotypic information related to metastatic potential that would enable the development of prognostic and predictive models. This application is enabled by radiomics, which can extract quantitative radiologic imaging features related to the aforesaid cellular phenotype, i.e., EMT-induced CSC changes, tumor budding, tumor islands, and STAS. [52] Furthermore, because of its potential correlation with pathologic morphology,[53-56] positron emission tomography (PET) has been studied to predict the pathologic outcome of therapy in various cancers, including lung,[57-60] cervix, [61,62] and other cancers.[63-65] These studies have revealed PET as a promising quantitative reflection of the pathologic heterogeneity at the tumor edges.

We developed the tumor shell, a radiomics feature that characterizes the tumor periphery and its correlation with distant failure. We demonstrated its ability in predicting treatment response for patients receiving SBRT for early stage NSCLC and for patients receiving EBRT and concurrent chemotherapy followed by high-dose-rate ICBT in stage IB-IVA CC.

## PATIENTS AND METHODS

### Patients

Our study was conducted at our institution, on two cohorts of patients approved by Institutional Review Board: (1) 48 early stage IA and IB NSCLC patients treated with SBRT from 2006 to 2012 (28 males and 20 females; mean age, 70.58 ± 9.84 years; range, 54 to 90 years); (2) 52 stage IB-IVA cervix cancer patients without para-aortic node involvement, treated with EBRT and concurrent chemotherapy followed by high-dose-rate ICBT from 2009 to 2012 (mean age, 47.10 ± 11.82 years; range, 26 to 72 years). In the NSCLC dataset, the total number of PET slices for each patient varied from 274 to 355, with 2.00 to 5.00-mm slice thickness and 4.0×4.0-mm or 5.0×5.0-mm pixel spatial resolution. Therefore, all slices were interpolated with the smallest slice thickness of 2.0 mm and spatial resolution of 4.0×4.0 mm to achieve a consistent format. In the CC dataset, all slices were used directly without interpolation since they had the same 5.00-mm slice thickness and 4.0×4.0-mm pixel spatial resolution. Before tumor analysis, the raw PET data were converted to standard uptake values (SUV).

### Tumor Analysis

For each patient, slices containing primary tumors were selected for analysis. In the NSCLC cohort, tumors were segmented automatically, with the middle location slice segmented by the object information based interactive segmentation method (OIIS) [66] and other slices segmented by the OTSU method.[67] In the CC cohort, the region of interest that incorporated the entire tumor was delineated manually by a radiation oncologist with 4 years' experience and reviewed by another radiation oncologist with 19 years' experience. In the NSCLC cohort, the number of selected slices originally ranged from 5 to 17, and zero padding was used for patients with slice numbers less than 17. Therefore, after interpolation to the smallest slice thickness of 2.0 mm, all patients had 42 slices. Meanwhile, because the greatest in-plane tumor diameter in all the patients' slices was 13 pixels, a patch of 17 × 17 pixels was cropped around the tumor center in each slice, resulting in a cube size of 17 × 17 × 42 for each patient. A volume size of 29 × 29 × 40 was used for each patient in the CC cohort. All features were computed on cropped PET cubes.

### Tumor Shell Feature Construction

The shell feature was extracted from the voxels around the tumor boundaries in a series of axial PET slices. The workflow of the shell feature construction is illustrated in Fig 1. The top row shows slices of the tumor (outlined in the red windows) in axial sequence (Fig 1). As displayed in the second row (Fig 1), the patches that include the delineated tumor were cropped from the corresponding slices above and used to compute the shell feature. By thresholding the patches above zero, binary mask images were obtained to represent the specific tumor region. By using the mask images of every two adjacent patches, a number of difference images were derived. As expected, a difference image was generally the outer region of the tumor. By adding up the difference images, referred to as sub-shells in the third row in Fig 1, the shell feature was formed and used to represent the holistic heterogeneity of voxels in the boundary of the entire tumor volume.

The sub-shell sequence $\Psi(t)$ is defined as:

$$\Psi(t) = \begin{cases} (M(t) - M(t-1)) \circ (P(t) - P(t-1)), & t \geq 2 \\ 0, & t = 1 \end{cases}$$
(1)

where $P(t)$ denotes a patch sequence and $M(t)$ is the matching binary mask image sequence that indicates the region of the tumor, the symbol ∘ indicates the hadamard product and $t$ is the patch (slice) number. When $t = 1$, $\Psi(t)$ is a zero matrix $\mathbf{0}$. The element in $\Psi(t)$ is either greater than (when corresponding elements in $M(t)$ and $M(t-1)$ are 0, 1 or 1, 0) or equal to zero (when corresponding elements in $M(t)$ and $M(t-1)$ are 0, 0 or 1, 1). Thus, each sub-shell partially describes the heterogeneous architecture of the tumor edge in an image where the higher SUV value pixels appear brighter. Examples of sub-shells are presented in Fig 1.

To represent the heterogeneity of the whole tumor border, for each patient $k$ the shell feature $S(k)$ is constructed by successively accumulating sub-shells together and can be written as:

$$S(k) = \sum_{t=2}^{n} \Psi(t). \qquad (2)$$

where $n$ is the total slice amount with $n = 42$ in the NSCLC cohort and $n = 40$ in the CC cohort. The strength of the shell feature is the use of a compact, yet comprehensive description that captures a sequence of morphologic patterns across the tumor boundary, such as shape, size, SUV values, and heterogeneities in a simple 2D map (Fig 1, bottom row).

### Handcrafted Feature

Our proposed shell feature was compared with the following five groups of handcrafted features: 9 intensity features, 8 geometry features, 12 second order gray level co-occurrence matrix (GLCM) features, 5 high order neighborhood gray tone difference matrix (NGTDM) texture features, and a combination of these four types, for a total of 34. The features are described in Table 2 and calculation functions are provided in the Supplement.

### Prediction Model Development

To develop our prediction model we used a machine learning method based on support vector machine (SVM). SVM is a supervised learning model that can classify data through an optimal hyperplane representing the largest separation margin between two classes. Before being fed to SVM, the vectorized shell was applied by principal component analysis (PCA) to reduce the feature dimension. The reduction process is described in the Supplement. The predictive ability of the shell feature was compared with that of the other five features using ten random trails of 5-fold cross validation on both NSCLC and CC cohorts. Meanwhile, to handle the class imbalance problem, SVM was trained over preprocessed data by a synthetic minority over-sampling technique (SMOTE). Accuracy, sensitivity, specificity, and area under the receiver operating characteristic curve (AUC) were used as evaluation metrics. The code was implemented in Matlab (version R2016a).

### Statistical Analysis

The difference in AUC performance between the shell feature and the other features was assessed by the Student's $t$-test. The difference was considered statistically significant with a $P$ value less than 0.05. The receiver operating characteristic (ROC) curve with 95% confidence interval is presented in Fig 2. Statistical analysis was performed with the Matlab statistical toolbox (version R2016a).

## RESULTS

### Clinical Characteristics

The demographic and clinical characteristics of patients in the NSCLC and CC cohorts are listed in Table 1. No significant difference in distant failure prevalence was observed between the two trials ($P = 0.917$). During follow-up time, distant metastases were observed in 25% (12 of 48) of patients in the NSCLC cohort and 26.9% (14 of 52) in the CC cohort after radiotherapy.

### Comparison of Predictive Performance

The comparison between the shell feature and other features was performed on both NSCLC and CC cohorts through quantitative analysis (Table 3) and ROC graphing (Fig 2). AUC, sensitivity, specificity, and accuracy were the criteria used in the study. Definitions are given in the Supplement.

The shell feature showed the highest accuracy in predicting distant failure (Table 3). In the NSCLC cohort, the shell feature achieved an AUC of 0.82 (95% CI, 0.6632 to 0.9247) with 0.81 sensitivity, 0.80 specificity, and 0.81 accuracy. For the other five features, the best result was observed for the GLCM texture as shown by 0.76 AUC (95% CI, 0.5528 to 0.8905), 0.75 sensitivity, 0.74 specificity, and 0.75 accuracy. Similarly, in the CC cohort the shell feature still achieved the best performance for all metrics, with 0.83 AUC (95% CI, 0.6559 to 0.9212), 0.81 sensitivity, 0.80 specificity, and 0.80 accuracy. These results revealed that the shell feature had more discriminative capacity than the other features. Also, the difference in AUC performance between the shell feature and the other features was found to be significant ($P < 0.005$ for both features in both cohorts).

The ROC curves for different feature sets are illustrated in Fig 2. Similar results were obtained for NSCLC (Fig 2A) and CC (Fig 2B). The proposed shell feature, represented by the upper blue curve, is located close to the top left corner of the chart, indicating on average a greater discriminative ability than the other methods.

The discriminative ability is indicated by representative 2D shell maps (Fig 3A and 3B) and vectorized shell feature matrixes (Fig 3C and 3D). The top row shows tumors without distant failure (Fig 3A) and the bottom row reports those with distant failure (Fig 3B). Pixels with higher SUV values are indicated in brighter colors, while lower values are shown in darker colors. As evident from the shell maps, distant failure-positive tumors show more heterogeneous boundary expression than the failure-negative ones. This finding may be attributed to the more active, varied, and potentially invasive cellular behavior of the tumor in the barrier microenvironment. The overall capability of the shell's classification for the NSCLC and CC cohorts is indicated in Fig 3C and 3D. The rows in the matrixes are the vectorized shell's sparse coefficients learned by the dictionary learning method.[68] Clustering characteristics can be seen on both cohorts, with features of the same class showing similar representation and features of different classes displaying distinct representations.

## DISCUSSION

The potential of tumor boundary as a predictive factor for distant failure was evaluated by the tumor shell, a PET-derived feature that allows us to detect its associations with metastasis within the microenvironment. The shell feature can be used to predict the outcome of SBRT for NSCLC patients and EBRT and concurrent chemotherapy followed with high-dose-rate ICBT for CC patients.

The tumor-host interface has been associated with metastasis because interactions between tumor cells and their microenvironment play an active part in tumor invasion and metastasis.[69] However, to the best of our knowledge, few studies have targeted tumor boundaries in medical imaging for constructing risk models of metastasis.[70,71] A recent study linked the morphology at the tumor-stroma interface to a multifractal metric, which derived from tumor outlines (excluding tumor internal tissue) on pathological images. The outline-based metric was found to be associated significantly ($P < 0.001$) with metastasis-related features, such as tumor border configuration and tumor budding grade, thereby verifying its prognostic and predictive efficacy for treatment response in colon cancer.[70] Similarly, in a lung cancer review, a fractal dimension of the tumor-stroma interface was used to measure tumor progression.[71] The study highlighted the use of radiological imaging, and found that the derived metric correlated with tumor growth and predicted treatment response.[71] Notably, the predictors in these studies were scores calculated from the contour lines of the tumor edge, whereas our method used the areas of the tumor boundary, where more minable information may be included. Besides, the calculation of the scores is a handcrafted processing, which is subject to human inconsistency and operator dependence. By contrast, our shell feature uses the original image information of the tumor edge directly and eliminates any calculation or feature selection procedure, thereby may be more scalable and generalizable.

The correlation between distant failure and radiomics features of the tumor edge is based on known biological processes that are associated with metastatic potential such as EMT and tumor budding. On the assumption that these findings are located at the tumor boundary, the shell feature was proposed to describe spatial morphology of the tumor periphery in relation to the likelihood of metastasis. Moreover, to the extent that these processes are present in other tumor types, it is likely that the shell feature may be used to predict the outcomes for other cancers.

Our study presents a few limitations, including the use of a small patient population and a 5-fold cross validation instead of an independent validation cohort. Also, accumulating a serial of sub-shells (3D) into a 2D shell feature may lead to a loss of spatial complexity in the axial perspective. Finally, the influence of tumor boundary extension is not investigated in this paper.

In conclusion, the PET-derived shell feature revealed a relationship between tumor edge and distant failure, and can be used to facilitate early prediction of the radiotherapeutic response in NSCLC and CC patients.

# TABLES & FIGURES

| Characteristics | NSCLC Cohort | | CC Cohort | |
|---|---|---|---|---|
| | Distant failure (+) | Distant failure (-) | Distant failure (+) | Distant failure (-) |
| Age, years | | | | |
|   mean ± SD | 69.9 ± 9.2 | 70.2 ± 10.2 | 41.6 ± 11.7 | 49.1 ± 11.3 |
|   Median (range) | 69.5 (57.0-89.0) | 71.5 (54.0-90.0) | 38.2 (29.3-70.9) | 49.1 (26.2-72.0) |
| Ethnicity, No. (%) | | | | |
|   Caucasian | 9 (75.0) | 27 (75.0) | 6 (42.9) | 12 (31.6) |
|   Hispanic | 0 (0) | 1 (1.3) | 2 (14.3) | 15 (39.5) |
|   African American | 3 (25.0) | 7 (19.4) | 5 (35.7) | 10 (26.3) |
|   Asian | 0 (0) | 1 (1.3) | 1 (7.1) | 0 (0) |
|   Other | 0 (0) | 0 (0) | 0 (0) | 1 (2.6) |
| Clinical tumor size, mm, No. (%) | | | | |
|   ≤ 10 | 1 (8.3) | 0 (0) | 0 (0) | 1 (2.6) |
|   11-30 | 6 (50.0) | 26 (72.2) | 2 (14.3) | 4 (10.6) |
|   31-50 | 5 (41.7) | 9 (25.0) | 9 (64.3) | 20 (52.6) |
|   51-70 | 0 (0) | 1 (1.3) | 2 (14.3) | 9 (23.6) |
|   > 71 | 0 (0) | 0 (0) | 1 (7.1) | 4 (10.6) |
| Histology, No. (%) | | | | |
|   Adenocarcinoma | 6 (50.0) | 17 (47.3) | 1 (7.1) | 4 (10.6) |
|   Squamous cell carcinoma | 5 (41.7) | 12 (33.3) | 11 (78.6) | 33 (86.8) |
|   Other | 1 (8.3) | 7 (19.4) | 2 (14.3) | 1 (2.6) |
| Stage, No. (%) | | | | |
|   IA | 5 (41.7) | 30 (83.3) | 0 (0) | 0 (0) |
|   IB | 7 (58.3) | 6 (16.7) | 4 (28.6) | 12 (31.6) |
|   IIA | 0 (0) | 0 (0) | 1 (7.1) | 3 (7.9) |
|   IIB | 0 (0) | 0 (0) | 7 (50.0) | 15 (39.5) |
|   IIIB | 0 (0) | 0 (0) | 0 (0) | 6 (15.8) |
|   IVA | 0 (0) | 0 (0) | 2 (14.3) | 2 (5.2) |

**Table 1.** Characteristics of two cohorts of patients

NOTE. Stages in NSCLC and CC are determined by the TNM and Federation of Gynecology and Obstetrics (FIGO) staging system, respectively.
Abbreviations: NSCLC, non-small cell lung cancer; CC: cervix cancer; SD: standard deviation.

**Table 2.** Types of handcrafted features

| Histogram based image intensity | Geometry | GLCM based texture | NGTDM based texture |
|---|---|---|---|
| Minimum | Volume | Energy | Coarseness |
| Maximum | Major diameter | Entropy | Contrast* |
| Mean | Minor diameter | Correlation | Busyness |
| Stand deviation | Eccentricity | Contrast* | Complexity |
| Sum | Elongation | Texture Variance | Texture Strength |
| Median | Orientation | Sum-Mean | |
| Skewness | Bounding Box Volume | Inertia | |
| Kurtosis | Perimeter | Cluster Shade | |
| Variance | | Cluster tendency | |
| | | Homogeneity | |
| | | Max-Probability | |
| | | Inverse Variance | |

NOTE. *Contrast: Different calculation methods were employed in GLCM and NGTDM, though the same names are indicated.
Abbreviations: GLCM, Gray level co-occurrence matrix; NGTDM, Neighborhood gray tone difference matrix.

| cohort | features | Accuracy | Sensitivity | Specificity | AUC | 95% CI | P value |
|---|---|---|---|---|---|---|---|
| | | | | | | | |
| NSCLC | Intensity | 0.70 ± 0.01 | 0.70 ± 0.02 | 0.69 ± 0.01 | 0.73 ± 0.02 | [0.5615,0.8613] | .0002 |
| | Geometry | 0.68 ± 0.01 | 0.65 ± 0.06 | 0.70 ± 0.04 | 0.65 ± 0.01 | [0.4861,0.8009] | .0001 |
| | GLCM texture | 0.75 ± 0.03 | 0.75 ± 0.03 | 0.74 ± 0.03 | 0.76 ± 0.02 | [0.5528,0.8905] | .0044 |
| | NGTDM texture | 0.68 ± 0.03 | 0.70 ± 0.06 | 0.65 ± 0.02 | 0.73 ± 0.03 | [0.5139,0.8783] | .0015 |
| | Combination | 0.73 ± 0.04 | 0.72 ± 0.02 | 0.71 ± 0.03 | 0.76 ± 0.02 | [0.5887,0.8796] | .0025 |
| | **Shell** | **0.81 ± 0.03** | **0.81 ± 0.02** | **0.80 ± 0.03** | **0.82 ± 0.03** | **[0.6632,0.9247]** | – |
| CC | Intensity | 0.72 ± 0.02 | 0.71 ± 0.03 | 0.75 ± 0.03 | 0.69 ± 0.01 | [0.4743,0.8533] | .0003 |
| | Geometry | 0.71 ± 0.04 | 0.71 ± 0.01 | 0.71 ± 0.04 | 0.71 ± 0.01 | [0.4891,0.8590] | .0006 |
| | GLCM texture | 0.75 ± 0.02 | 0.80 ± 0.02 | 0.73 ± 0.02 | 0.76 ± 0.04 | [0.5427,0.8981] | .0015 |
| | NGTDM texture | 0.72 ± 0.04 | 0.71 ± 0.02 | 0.74 ± 0.04 | 0.74 ± 0.03 | [0.5396,0.8524] | .0002 |
| | Combination | 0.72 ± 0.03 | 0.75 ± 0.05 | 0.73 ± 0.03 | 0.73 ± 0.02 | [0.5519,0.8813] | < .0001 |
| | **Shell** | **0.80 ± 0.04** | **0.81 ± 0.02** | **0.80 ± 0.04** | **0.83 ± 0.02** | **[0.6559,0.9212]** | – |

**Table 3.** Prediction performance of features with respect to distant failure

NOTE. "Combination" refers to the combined four types of features, i.e., intensity, geometry, GLCM texture and NGTDM texture; 95% CI and *P* value are both derived from values of AUC; *P* value measures the statistical AUC difference between each group of handcrafted features and shell feature.

Abbreviations: AUC, the area under a characteristic operation curve; CI, confidence interval.

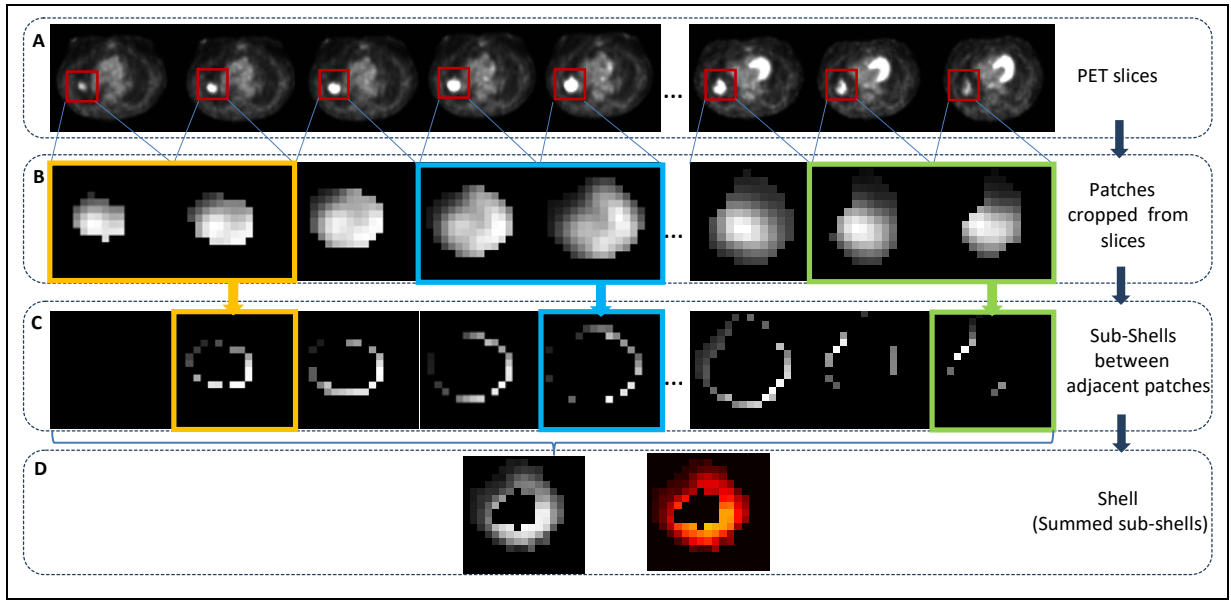

**Fig 1.** Shell feature extraction workflow. (A) Series of axial PET slices of one patient. (B) Series of patches (red windows in A) including tumors are cropped from each slice in A. (C). Series of sub-shells derived from adjacent two patches in B. (D). Shell feature, with grayscale image left and Heatmap image right.

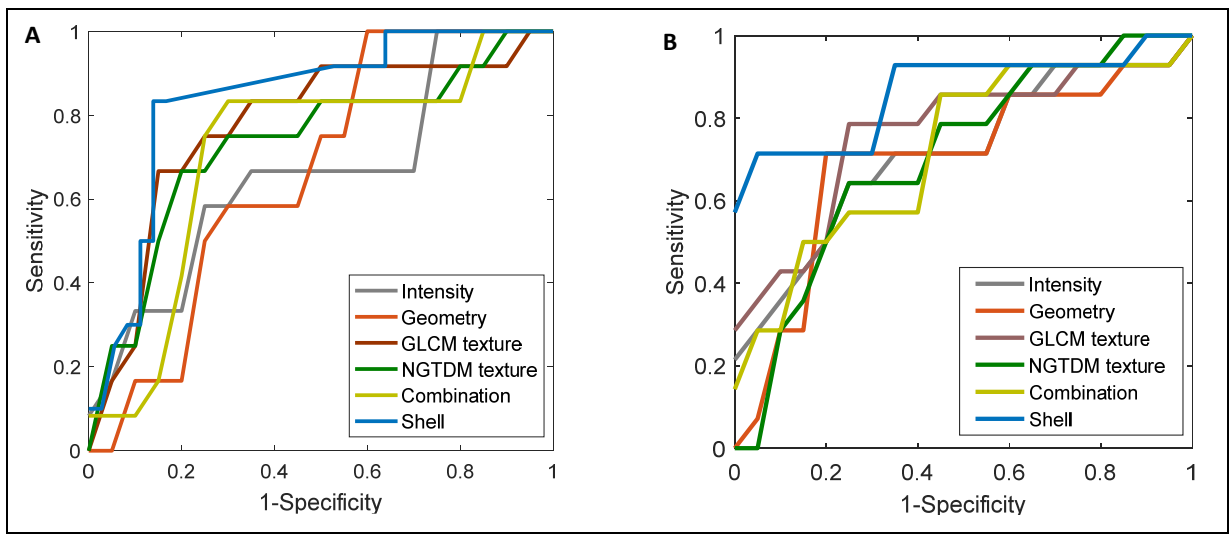

**Fig 2.** Receiver operating characteristic (ROC) curves of shell feature and other five groups of handcrafted features. (A) NSCLC cohort. (B)CC cohort. ROC curves depict the classification ability of the binary SVM model in terms of predictive feature and observed outcome of distant failure under varied discrimination threshold. The *x*-axis represents the false positive rate and is calculated as (1-specificity). The *y*-axis represents the true positive rate by sensitivity. A larger area under the curve indicates better prediction.

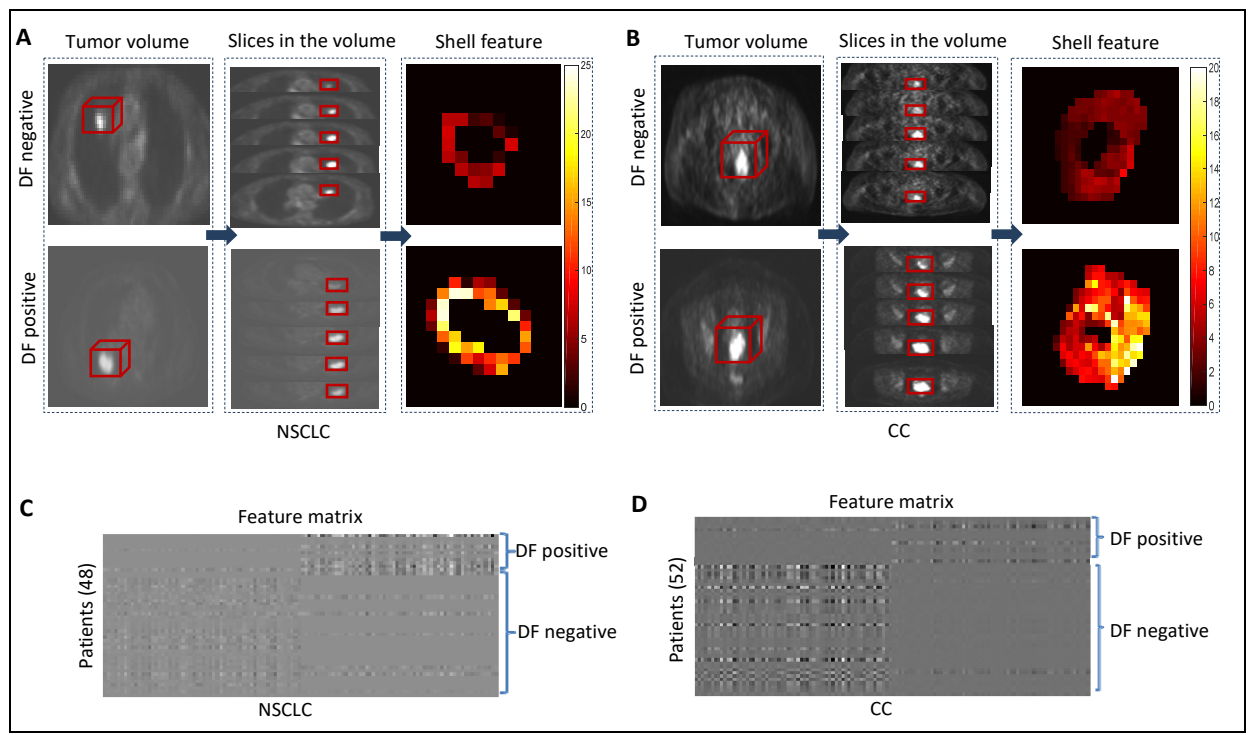

**Fig 3.** The shell feature has the discriminative ability to detect distant failure (DF)-negative and -positive tumors. (A) (B) are representative examples of 2D shell in terms of structure heterogeneity. (A) NSCLC cases. (B) CC cases. In each cohort, the shell feature (third column) is computed from a series of slices (second column) in the tumor volume (first column), with the top row showing tumors without distant failure and the bottom row showing tumors with distant failure. As shown, tumors with distant failure present more complicated morphologic patterns. (C) (D) are feature matrixes of the whole patients, where each row corresponds to a patient, and each column corresponds to an element of the feature. (C) NSCLC cases. (D) CC cases. These features are sparse coefficients learned from the original vectorized shells by dictionary learning method. These feature matrixes exhibit clustering characteristics for (DF)-positive and -negative tumors.